# Investigating radiation damage in nuclear energy materials using JANNuS multiple ion beams

A. Gentils[1], C. Cabet[2]

[1] CSNSM, Univ Paris-Sud, CNRS/IN2P3, Université Paris-Saclay, 91405 Orsay, France
[2] DEN, Service de Recherches de Métallurgie Physique, CEA, Université Paris-Saclay, 91191 Gif sur Yvette, France

**Corresponding authors**:
A. Gentils (aurelie.gentils@u-psud.fr), C. Cabet (celine.cabet@cea.fr)

**Abstract**: Ion accelerators have been used by material scientists for decades to investigate radiation damage formation in nuclear materials and thus to emulate neutron-induced changes. The versatility of conditions in terms of particle energy, dose rate, fluence, etc., is a key asset of ion beams allowing for fully instrumented analytical studies. In addition, very short irradiation times and handling of non-radioactive samples dramatically curtail the global cost and duration as compared to in-reactor testing. Coupling of two or more beams, use of heated/cooled sample holders, and implementation of *in situ* characterization and microscopy pave the way to real time observation of microstructural and property evolution in various extreme radiation conditions more closely mimicking the nuclear environments. For these reasons, multiple ion beam facilities have been commissioned worldwide.

In France, under the auspices of the Université Paris-Saclay, the JANNuS platform for 'Joint Accelerators for Nanosciences and Nuclear Simulation' comprises five ion implanter and electrostatic accelerators with complementary performances. At CSNSM (CNRS & Univ Paris-Sud, Orsay), a 200 kV Transmission Electron Microscope is coupled to an accelerator and an implanter for *in situ* observation of microstructure modifications induced by ion beams in a material, making important contribution to the understanding of physical phenomena at the nanoscale. At CEA Paris-Saclay, the unique triple beam facility in Europe allows the simultaneous irradiation with heavy ions (like Fe, W) for nuclear recoil damage and implantation of a large array of ions including gasses for well-controlled modelling-oriented experiments.

Several classes of materials are of interest for the nuclear industry ranging from metals and alloys, to oxides or glasses and carbides. This paper gives selected examples that illustrate the use of JANNuS ion beams in investigating the radiation resistance of structural materials for today's and tomorrow's nuclear reactors.

## 1. Introduction

Under the auspices of the new Université Paris-Saclay, the multi-ion beam irradiation platform JANNuS for 'Joint Accelerators for Nanosciences and Nuclear Simulation' is dedicated to researches in the effects of ions in materials. Its particle beams make it possible to irradiate small samples in a perfectly controlled manner, and thus to observe and quantify the evolution of their microstructure (segregation, precipitation, formation of dislocation loops, cavities, bubbles, etc.) and service properties. JANNuS-Orsay and JANNuS-Saclay have been developed from the origin as two complementary facilities on neighbouring sites (Orsay and Saclay) and they are bound since 2005 by a Grouping of Scientific Interest (GIS) [1]. This platform offers to the international academic community the opportunity to perform fully instrumented irradiation experiments on advanced materials which may be supplemented by *in situ* characterization techniques such as Transmission Electron Microscopy, Raman or Ion Beam Analysis (IBA) [2]. The actual capabilities combine five electrostatic accelerators for single beam, dual beam and triple beam ion experiments and a Transmission Electron Microscope (TEM) for *in situ* studies. Such a scientific platform has no equivalent in Europe and plays an essential role for multi-scale modelling of radiation effects in materials. With the scarcity of Material Testing Reactors throughout the world, ion irradiation facilities allow to emulate neutron irradiation and to understand the fundamentals of radiation resistance of nuclear materials. This is of special importance for the present and future nuclear industry. As an example, Ref. [3,4] describe the need to predict the modification of materials under irradiation and to validate their radiation resistance.

*In situ* Transmission Electron Microscopy is a speciality since the early 1980's [5] of the CSNSM lab (joint research unit of CNRS/IN2P3 and Université Paris-Sud) located in Orsay, France. A 120 kV Philips EM400 TEM and a 190 kV homemade ion implanter (called IRMA [6]) were connected together under the guidance of Dr. Marie-Odile Ruault, allowing *in situ* observations of modification of materials under ion beam. Several research projects took place mainly on ion beam synthesis in semiconductors and metals using this peculiar equipment [e.g. 7-10]. A new 120 keV Philips CM12 microscope equipped with Energy-Dispersive X-ray Spectroscopy was installed in 1994 [11] and researches were still focused on ion beam synthesis in silicon for microelectronic applications [e.g. 12], but also dedicated to nuclear materials [e.g. 13-15]. The facility was updated in 2006 [16] with the arrival of a new FEI Tecnai 200kV G$^2$20 TEM and the construction of a new ion beam line connected to ARAMIS [17], a 2 MV Tandem / Van de Graaff homemade accelerator, built in the late 80's. This exceptional facility that includes the TEM and the two ion beam lines coming from the IRMA ion implanter and the ARAMIS ion accelerator has been called JANNuS-Orsay. It is part of the SCALP accelerators platform of the CSNSM lab, where single ion implantation/irradiation and ion beam analysis are also available [18].

The DMN department of the CEA Paris-Saclay has been developing for many years modeling tools from the atomic to the macroscopic scale in order to validate the resistance of nuclear materials under the extremely harsh conditions which they encounter in service, in particular irradiation effects, and to design innovative materials for advanced nuclear systems [19]. These models rely on high-performance numerical methods and on fully controlled characterization techniques at the same scales. In line with this technological and scientific approach, the SRMP has designed the JANNuS-Saclay triple ion beam irradiation facility as a key tool to understand the physical mechanisms of neutron radiation damage, and to validate the multiscale modeling of the macroscopic events of materials aging [20].

After a brief technical description of each facility, this paper gives selected examples of recent research studies related to nuclear materials for existing and future reactors that have been performed at JANNuS.

## 2. Description of the equipments

Ion accelerators have been used by material scientists for decades to investigate radiation damage formation in nuclear materials and thus to emulate neutron-induced changes [e.g. 21, 22]. The versatility of conditions in terms of particle energy, dose rate, fluence, is a key asset of ion beams allowing for fully instrumented analytical studies. In addition, very short irradiation times and handling of non-radioactive samples dramatically curtail the global cost and duration as compared to in-reactor testing. Coupling of two or more beams, use of heated/cooled sample holders, and implementation of *in situ* characterization and microscopy pave the way to real time observation of microstructural and property evolution in various extreme radiation conditions more closely mimicking the nuclear environments. For these reasons, multiple ion beam facilities have been commissioned worldwide. A few triple ion beam facilities are operating at Takasaki Advanced Radiation Research Institute (TIARA), Japan [23], at the National Tsing Hua University (Accelerator Laboratory) in Taiwan [24], or at the University of Michigan (MIBL) in USA [25], to name only a couple. *In situ* TEM with ions (single or dual beam) facilities are available at various locations, such as in the USA, Japan, UK, China, as described in Ref. [26].

In France, within the Université Paris-Saclay, the JANNuS platform comprises five ion implanter and accelerators with complementary performances presented below, that make it quite unique in the world. A more detailed description of the facilities can be found in Ref. [27].

### 2.1. **JANNuS-Orsay** *in situ* **dual ion beam TEM**

The JANNuS-Orsay facility, a scheme of which is shown in Figure 1, is now operating according to three modes: i) TEM + IRMA ion implanter, ii) TEM + ARAMIS ion accelerator and iii) TEM in dual ion beam mode (TEM + IRMA + ARAMIS), at a chosen temperature in the range 77 - 1300 K, allowing *in situ* observation and analysis of the material microstructure modifications induced by single or dual ion implantation/irradiation. The detailed characteristics of IRMA ion implanter and ARAMIS Van de Graaff-Tandem ion accelerator are fully described in Ref. [6, 17, 18]. More than 40 elements from H to Yb can be produced from their ion sources, and both beams are rastered during ion implantation/irradiation, so that ion irradiation of the observed zone is homogeneous. Inside the TEM, the typical range of beam energies available depends on elements and is within 10-500 keV for the IRMA 190 kV ion implanter and 0.5-6 MeV for the 2 MV ARAMIS ion accelerator. Indeed due to the deflection of the ion beam going to the TEM, a limitation of a maximum energy of 1 MeV per charge state occurs in the microscope for ions coming from ARAMIS. *In situ* simultaneous dynamical TEM observation is possible when using one or two ion beam lines, depending on the geometry used (i.e. tilt angle values, shape of the thin foil, location of the transparent area, use of a ultra-thin specimen holder to minimize shadowing effects, etc.). If using light and low energy elements (below 100 keV N ions) on the IRMA ion beam line, only sequential observation is possible due to the ion beam deflection induced by the magnetic field of the objective lens of the microscope, that must be switched off during *in situ* ion implantation of such elements. For each ion beam line an open Faraday cup is located close to the TEM thin foil less than 3 cm away (thanks to the modified polar pieces), so that the current is measured continuously during the ion irradiation. The measurements of the flux and fluence are given with an accuracy of 10 %. The typical flux range measured in the microscope, depending on elements and energies, is between approximately $1 \times 10^9$ cm$^{-2}$.s$^{-1}$ and $5 \times 10^{11}$ cm$^{-2}$.s$^{-1}$. The two ion beam lines have a 45° angle between them, and 22° with respect to the surface normal direction, as shown in the inset in Figure 1. The TEM is a 200 kV FEI Tecnai G$^2$ 20 Twin equipped with a LaB$_6$ filament, with a spatial resolution of 0.25 nm. Images and videos are recorded using a 2kx2k CCD high-resolution camera, with 30 frames per second recording, or a high speed and wide area-imaging camera. Several analytical techniques are also coupled to the microscope, e.g Electron Energy Loss Spectroscopy (EELS), Energy-Filtered TEM (EFTEM), Scanning TEM (STEM) and Energy-Dispersive X-ray Spectroscopy (EDXS), allowing for example a chemical analysis of the specimen.

## 2.2 JANNuS-Saclay triple ion beam

The JANNuS-Saclay facility consists of three electrostatic accelerators, respectively named Épiméthée, Japet and Pandore, connected to a triple beam chamber for single-, dual- and triple beam irradiations [28]. Two other chambers are linked to Épiméthée and Pandore for single beam irradiation and/or Ion Beam Analysis, as shown in Figure 2.

Épiméthée is 3 MV, single-ended (NEC, 3-UH-2 Pelletron) and equipped with an electron cyclotron resonance (ECR) source (Pantechnik, Nanogan type). The ECR source delivers high charge states of gas ions like O, He, Ar, Xe, H and metal ions such as Fe and W, which are produced by the MIVOC process from two cartridges of organometallic compounds. In combination with the 3 MV acceleration potential, it can supply ion beams with energies from 0.5 MeV up to as high as 36 MeV. Maximal current at the targets is for example 140 µA for protons and He$^+$, 12 µA for Fe$^{5+}$, 7 µA for W$^{5+}$. Épiméthée is thus able to provide rather high dose rates of Fe and W that are classically used to induce damage levels up to 1 – 2 µm range in metals with damage rate typically up to 10 dpa/h.

Japet is a 2 MV Tandem (NEC, 6SDH-2 Pelletron) equipped with an external Cs sputtering source (SNICS II) which can deliver a large variety of ions such as Cl, I, C, Si, V, Cu, Zr, Ag, Au… Associated to an analysis magnet at 90°, this system produces beams protected from any contamination with energies of 0.5 to about 14 MeV depending on the charge state. Typical beams and current include: protons at 50 µA, iodine at 4 µA and gold at 1 µA.

Pandore is 2.5 MV, single-ended and equipped with a RF source which produces single-charge gas ions like protons, deuterium ions, helium-4 and helium-3 ions. This accelerator is used for implantation as well as for IBA. Maximal current intensities are for example 2 µA for protons and 9 µA for $^4$He.

The triple beam chamber receives one beam line coming from each accelerator with an incidence angle of 15 ° allowing single, dual or triple beam irradiations. A second vacuum chamber in line with Épiméthée can be used for single beam irradiation at normal incidence. Beam lines can be raster scanned to spread the beam homogeneously over an area of 2 cm by 2 cm on the target and they can be complemented with energy degraders that flatten irradiation damage and implanted gas in the sample thickness. During irradiation experiment, current intensity is monitored using a mobile multi-pin Faraday cups device; integrating current intensity provides an accurate quantification of implanted species. Pumping groups and cold traps allow reaching a vacuum better than $10^{-7}$ Torr. Heating-cooling sample stages provide precise temperature control from liquid nitrogen temperature to 800°C. One to five thermocouples are used to monitor the sample temperature and a 2D infrared thermal imaging camera maps the sample surface temperature. At higher temperature, a bi-chromatic pyrometer can be implemented. A confocal Raman spectrometer (Renishaw, Invia Reflex) has been connected to the triple beam chamber in 2014 for *in situ* measurements [29]. It has been successfully used for investigating the kinetics of radiation damage in various systems.

## 3. Examples of radiation damage investigations in nuclear materials at JANNuS

In the last decade, many materials of interest for the nuclear industry have been investigated at the JANNuS platform. A non-exhaustive list includes: nuclear fuel and surrogates such as uranium dioxide [30-33]; neutron absorber like $B_4C$ [34,35]; structural alloys for present, advanced and future fission and fusion systems as well as model metals such as iron, Fe-Cr alloys and ferritic-martensitic steels [36-39], ODS steels [40-45], austenitic steels [46-51], tungsten and copper [52-54], aluminum alloys [55], zirconium [56-58]; Enhanced Accident Tolerant Fuel [59, 60]; advanced structural ceramics for innovative nuclear systems such as carbides and nitrides [61-65]; glasses and oxides for waste management [66-69]; advanced coatings for Liquid Metal fast Reactors [70]; polymers for cable sheath or glove box components [71]. Selected studies are presented in the following that exemplify the prime role of *in situ* characterization during ion irradiation experiments and the strength of associating numerical modelling and controlled ion irradiation. All these examples are aimed to increase the understanding of fundamental properties in materials submitted to irradiation, for actual and future nuclear industries [72,73].

### 3.1. Stability of ODS nano-oxide particles under ion irradiation

Oxide Dispersion Strengthened (ODS) steels are considered as promising candidates for structural material in future fission and fusion reactors. These steels exhibit enhanced HT creep and mechanical strength thanks to the dispersion of nano-sized oxide particles.

Different experiments were performed to confirm the stability of nano-oxides under ion irradiation in ferritic-martensitic ODS steels [40-44], and in austenitic ODS steels [45]. The stability of nano-oxides in a ODS Fe-18Cr stabilized with $Y_2O_3$ was investigated by *in situ* TEM at JANNuS-Orsay under Fe-ion irradiation. It was shown that Y-Ti-O nanoprecipitates as small as 5 nm are stable up to 45 dpa (displacements per atom), and larger oxides seem more affected by ion irradiation [40]. Irradiation at higher doses allowed studying the nanoparticle evolution [40]. At room temperature, irradiation at 200 dpa at JANNuS-Orsay lead to a complete dissolution of the oxide particles. TEM observations and APT analysis of samples irradiated at 500°C and 150 dpa at JANNuS-Saclay revealed an increase in particle size and a decrease of their density with increasing irradiation dose. It was assumed that the most important effect of irradiation at 500 °C would be to accelerate kinetics and therefore to favour a faster evolution towards thermodynamic equilibrium.

With the aim of clarifying the role of nanoparticles, single (Fe)-, dual (He)- and triple (H)-beam ion irradiations were also performed in ferritic/martensitic steels w/o ODS reinforcement and microstructural damage was characterized by TEM [42, 43]. Dislocation loops were observed in Eurofer-97 and Eurofer-ODS steels under single Fe irradiation as well as under dual and triple beam irradiation at 26 dpa and 400 °C [42]. However, simultaneous gas ion implantation induces the formation of cavities. It was shown that a high-density of ODS nano-particles such as in the ferrite grains of Eurofer-ODS inhibits cavity formation.

### 3.2. Boron carbide behaviour under ion irradiation

Boron carbide is widely used as a neutron absorber in nuclear plants. $B_4C$ samples were irradiated with Au ions at different temperatures at JANNuS and damage evolution was monitored by *in situ* TEM [34] and *in situ* Raman [35]; it should be noted that $B_4C$ exhibits characteristic Raman fingerprints for the pristine as well as for the irradiated structure that enable its characterization by Raman spectroscopy. As an example, *in situ* Raman spectroscopy was performed during 4 MeV Au ion irradiations up to $10^{15}$ cm$^{-2}$ at -160°C, room temperature, and 800°C. The Raman analyses showed a structural disorder starting at a low fluence of $10^{13}$ cm$^{-2}$, equivalent to ~0.01 dpa. However, post-irradiation TEM examinations have shown that the structure of the material remains crystalline. At a higher fluence, if one keeps away from the Au-ion implanted zone, small amorphous areas of few nanometers appear in the damaged zone but the long-range order is preserved. The material is fully amorphous above about 7.5 dpa at room temperature but at a higher dpa level at 800°C. Figure 3 shows the Raman spectrum evolution at room temperature. The defect band (marked by an arrow) is attributed to a loss of symmetry. This band gets broader from $9x10^{11}$ to $9x10^{14}$ cm$^{-2}$, and then it remains almost constant in width. These graphs illustrate that *in situ* Raman may be a powerful tool to monitor damage kinetics during irradiation.

### 3.3. Ageing of austenitic steels under irradiation

Single and dual ion beam irradiation experiments were used to study the ageing of austenitic stainless steels under radiation w/o helium gas presence [46-51]. The synergistic effects of the defect creation (induced by 4 MeV $Au^{2+}$) and He injection (10 keV $He^+$) on their potential swelling were for example studied using *in situ* dual ion beam TEM observations at JANNuS-Orsay [47]. Characterization of the populations of cavities and defects induced by single or dual ion beam was performed at different temperatures and fluences, for both industrial CW 316L steel and model FeNiCr alloy. The importance of the minor constituents in the microstructure under ion irradiation was highlighted. Figure 4 shows the cavities microstructure (in dark grey/black) obtained after single Au ion irradiation (a), consecutive single Au ion irradiation and He implantation (b), and dual ion beam irradiation (c). It is observed that simultaneous He implantation enhances the cavity nucleation in contrast to when He is injected after damage. Moreover it was shown that the cavity distribution characteristics agree qualitatively with rate theory models.

### 4. Conclusion

In France, under the auspices of the Université Paris-Saclay, the JANNuS platform for Joint Accelerators for Nanosciences and Nuclear Simulation comprises five ion implanter and electrostatic accelerators with complementary performances. At CSNSM (CNRS/IN2P3 and Univ Paris-Sud, Orsay), a 200 kV Transmission Electron Microscope is coupled to an accelerator and an implanter for *in situ* observation of microstructure modifications induced by ion beams in a material, making important contribution to the understanding of physical phenomena in irradiated materials at the nanoscale, with a wide range of ions and energies available. At CEA Paris-Saclay, the unique triple beam facility in Europe allows the simultaneous irradiation with heavy ions (like Fe, W) for nuclear recoil damage and implantation of a large array of ions including gasses for well-controlled modelling-oriented experiments. Some current investigations on various systems of interest for the nuclear industry have been briefly described in this paper and complementary information can be found in the list of references. These fundamental researches allow better understanding of the physical mechanisms that occurred in nuclear materials submitted to extreme conditions such as irradiation, gas presence and temperature. This is significant for materials currently used in the nuclear industry as well as for candidates that may be used in the future.

The JANNuS platform is open to the international academic community for scientifically relevant proposals through the call for projects of the EMIR French accelerators federation [74]. Our wish for the near future is to keep up with the state-of-the-art ion beam technologies and to reinforce *in situ* characterization to monitor the microstructure modification of materials under ion beams.

### Acknowledgments

The current and former JANNuS staffs are gratefully acknowledged for their unfailing assistance. The on-going support is provided at CSNSM Orsay especially by C. Bachelet, C.

**Figures Captions**

**Figure 1**. Overview of the JANNuS-Orsay facility, showing the coupling between the Transmission Electron Microscope and the two ion accelerators. The inset shows the geometry of the *in situ* TEM experiment with two ion beams.

**Figure 2**. Artist view of the JANNuS-Saclay facility.

**Figure 3**. *In situ* Raman spectroscopy of $B_4C$ irradiated with 4 MeV Au ions at room temperature. (a) Raman spectra at an increasing fluence (values indicated in $cm^{-2}$) and (b) evolution of the defect band width (see arrow in the graph (a)).

**Figure 4**. Overfocused bright field TEM images of 316L thin foils irradiated at 550°C (a) with Au ions to the fluence of $1x10^{15}$ $cm^{-2}$, (b) with Au ions to the same fluence, and subsequently implanted with He ions to the fluence of $1x10^{15}cm^{-2}$, (c) simultaneously Au irradiated to $1x10^{15}$ $cm^{-2}$ and He implanted to $3.5x10^{14}$ $cm^{-2}$, showing cavities (in black) induced by irradiation (modified from [47]).

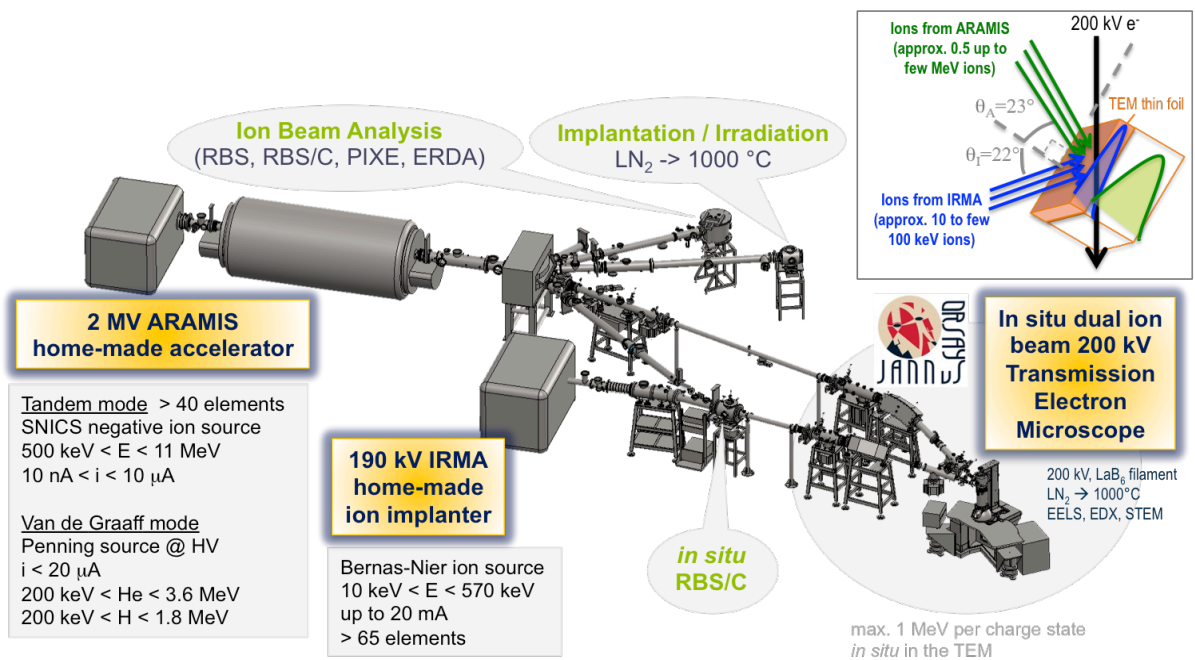

**Figure 1**
2-column figure
color on-line

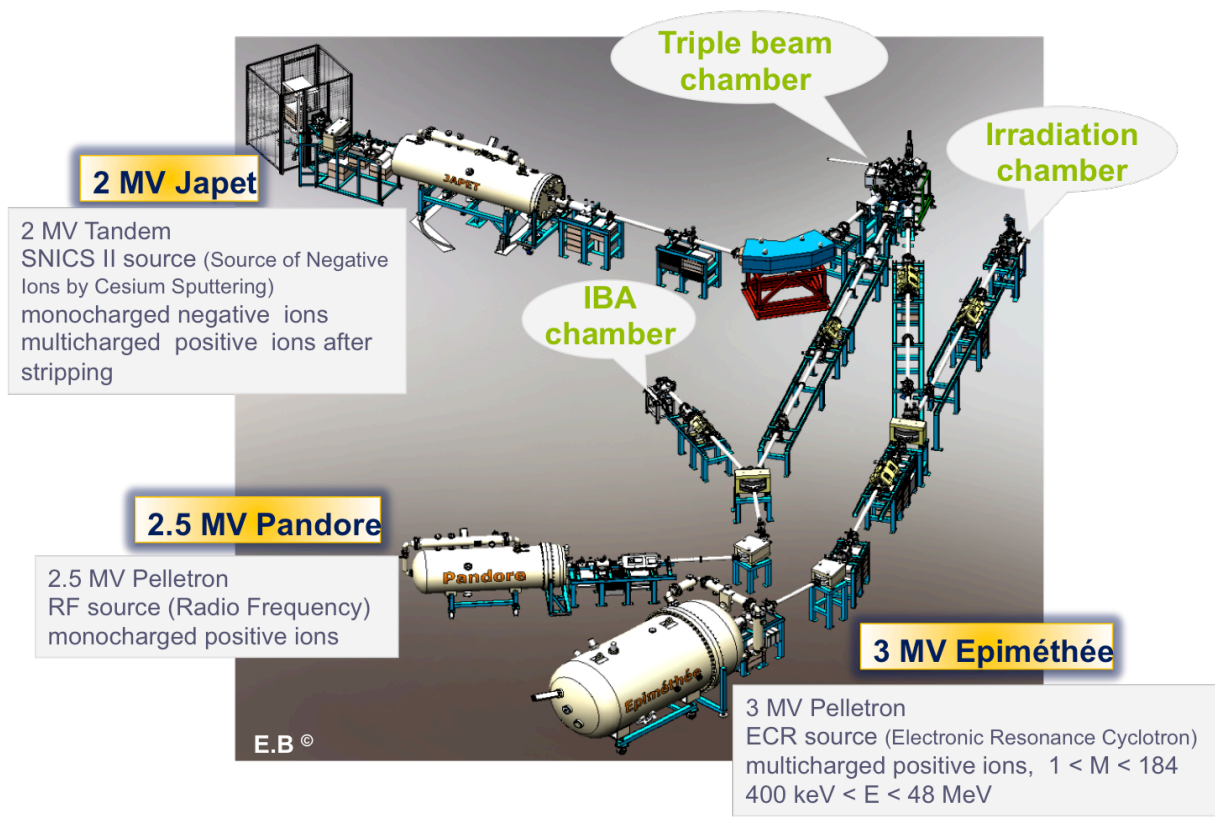

**Figure 2**
2-column figure
color on-line

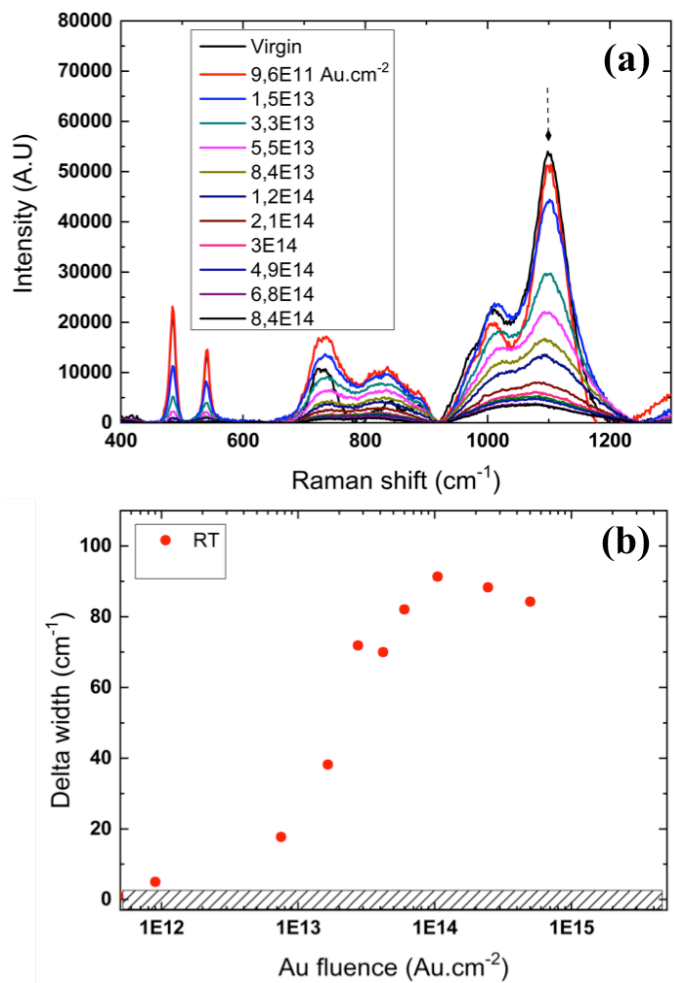

**Figure 3**
1-column figure

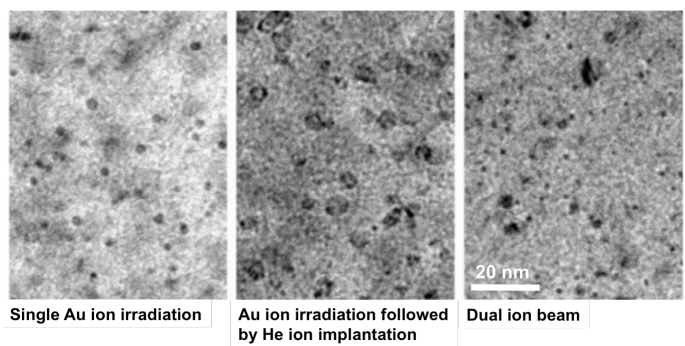

Single Au ion irradiation | Au ion irradiation followed by He ion implantation | Dual ion beam

**Figure 4**
1-column figure